\documentstyle[12pt]{article}

\begin{document}
\baselineskip=18pt plus 0.2pt minus 0.1pt
\parskip = 6pt plus 2pt minus 1pt
\newcommand{\reseteqnum}{\setcounter{equation}{0}}
\renewcommand{\theequation}{\thesection.\arabic{equation}}
\newcommand{\bx}{\mbox{\boldmath $x$}}
\newcommand{\bX}{\mbox{\boldmath $X$}}
\newcommand{\bY}{\mbox{\boldmath $Y$}}
\newcommand{\rhovev}{(\rho v)^2}
\newcommand{\mr}{\frac{g^2}{\lambda}}
\newcommand{\bra}[1]{\left\langle #1 \right|}
\newcommand{\ket}[1]{\left| #1 \right\rangle}
\newcommand{\VEV}[1]{\left\langle #1 \right\rangle}
\newcommand{\braket}[2]{\VEV{#1 | #2}}
\newcommand{\calP}{{\cal P}}

\input epsf

\begin{titlepage}
\title{
\hfill
\parbox{4cm}{\normalsize KUNS-1391\\HE(TH)~96/04\\YITP 97-11\\hep-ph/9703457}\\
\vspace{1cm}
Electroweak baryon and lepton number \\violating processes at high
energies \\ by the valley instanton  }
\author{Toshiyuki Harano
\thanks{e-mail address: \tt harano@gauge.scphys.kyoto-u.ac.jp}
\vspace*{0.5ex}
\\
{\normalsize\sl Department of Physics, Kyoto University, Kyoto 606-01,
  Japan}
\vspace*{1ex}
\\
and
\vspace*{1ex}
\\
Masatoshi Sato
\thanks{e-mail address: \tt msato@yukawa.kyoto-u.ac.jp}
\thanks{Address after April 1, 1997:
{\it Physics Department, University of Tokyo, Tokyo 113, Japan}}
\vspace*{0.5ex}
\\
{\normalsize\sl Yukawa Institute for Theoretical Physics,
Kyoto University, Kyoto 606-01, Japan}\\
}
\date{\normalsize March 1997}
\maketitle
\thispagestyle{empty}
\begin{abstract}
\normalsize
We calculate the cross sections of the baryon and lepton number violating
processes. The proper valley method enables us to calculate
the multi-boson processes, which have the possibility to observation.
The mass corrections of the gauge boson, Higgs particle and top quark are
evaluated and the group integration and the phase space integration
are performed analytically and numerically.
\end{abstract}
\end{titlepage}

\newpage
\reseteqnum
\section{Introduction}
It is well known that the baryon and lepton numbers are not exactly
conserved in the standard model, due to the chiral anomaly.
The topologically non-trivial configuration, especially, instanton
induces the baryon and lepton number violating process as the
following;
\begin{eqnarray}
{\rm q}+{\rm q}\rightarrow (3n_f-2)\bar{{\rm q}}+n_f\bar{l}+n_w {\rm
W}+n_h {\rm H},
\label{process}
\end{eqnarray}
where W represents gauge
boson, ${\rm W}^{\pm}$, Z and H represents the Higgs particle.
We denote the number of generations by $n_f$ and the numbers of the
bosons by $n_w$, $n_h$.
As 't Hooft first indicated\cite{thooft}, this process is highly
suppressed by the factor
$\exp(-16\pi^2/g^2)$ at low energies. This is because the process is
accompanied by quantum tunneling through the potential barrier that
separates the topologically distinct vacua. The top of the potential,
which is a saddle point in the functional space, corresponds to the
static unstable solution, so-called sphaleron\cite{manton}, and the
height of the potential barrier is given by the energy of the
sphaleron: $E_{sph}\sim m_{w}/\alpha$. In naive picture, the
tunneling rate grows up as the energy of the initial state becomes
higher, and it enhances when the initial state has high energies
comparable to $E_{sph}$. Since $E_{sph}$ is O(10) TeV,
the suppression is expected to decrease or vanish and the baryon and lepton
number violating processes be observable in TeV region.

In this
viewpoint, the high energy behavior of the
process (\ref{process}) has been studied in recent years\cite{RS}. By the
sphaleron-based calculation\cite{AG}, it was suggested that the scattering
processes involved with many gauge bosons and Higgs particles are not
exponentially suppressed.
Ringwald\cite{Ring} and Espinosa\cite{Esp} calculated the baryon
and lepton number violating scattering amplitudes semi-quantitatively
by the constrained instanton method\cite{Aff}, and they also suggested
that the amplitudes of the multi-boson
process increase with the energy and reach 1pb for a center of mass
energy of O(10) TeV.
Unfortunately, the high-energy behavior of the process with many bosons,
which is most interesting, is out of the validity of their calculations.
One of the problem is that they used the
constrained instanton.
The constrained instanton method is not available for the
large-radius instanton, which plays an important role in multi-boson
process. It is pointed out in Ref.\cite{AHSW} that the approximation
breaks for $n_w+n_h\sim 4$ in the constrained instanton method.

Another problem of Ref.\cite{Ring} is that the
orientation effect of the instanton is not taken into account. The
integration of the orientation in the Lorentz
SU$(2)_R$ rotations is not trivial but very important especially for
processes with many gauge bosons.
Moreover, in considering the multi-boson
processes at high energies, masses of the gauge boson, Higgs particle
and top quark cannot be neglected. All the effect remains unevaluated.

In this paper, we overcome these problems and evaluate the cross
section.
Instead of the constrained instanton method, we use the proper valley
method\cite{AKnv}
, which enables us to calculate the multi-boson
process for $n_w+n_h\sim 40$ \cite{AHSW}.
The integrations of the orientation and
phase space integration are performed analytically in the extremely
relativistic limit and non-relativistic limit, and performed
numerically in the general cases. The mass corrections of the gauge
boson, Higgs particle and top quark are taken into account.



In Section 2, we consider the valley instanton, which is constructed
by the proper valley method. This configuration gives an important
contribution to the path integral in topologically non-trivial sector
of the functional space.
In Section 3, we introduce the collective coordinates for the valley
instanton.
In Section 4,
we calculate the multi-point Green function $
\langle\psi^{4n_f}A_{\mu}^{n_w}H^{n_h}\rangle$ by the proper valley
method.
In performing the group integration and phase space integration, we
obtain the cross sections of the baryon and lepton number violating
processes.
In Section 5,  we give conclusion and discussion.

\reseteqnum
\section{Valley instanton in the electroweak theory}
\newcommand{\lsim}{\,\lower0.9ex\hbox{$\stackrel{\displaystyle
      >}{\sim}$}\,}
\newcommand{\rsim}{\,\lower0.9ex\hbox{$\stackrel{\displaystyle
      <}{\sim}$}\,}
We consider the standard model with $n_f$ generations of quarks
and leptons.
For simplicity, we set ${\rm SU(3)}_c$ and
${\rm U(1)}_Y$ coupling constants to zero, and set the
Kobayashi-Maskawa matrix to the identity.
Thus we treat the
simplified standard model that reduces to an SU(2) gauge-Higgs system
with fermions\cite{Esp,AHSW}. As for quarks, we have the left-handed doublets
$q_L^{i a}$ and the right-handed singlets $q_R^{i a}$, where
$a=1,2,3$ is the color index and $i=1,\cdots,n_f$
is the index of the generations. Similarly, we have the left-handed
doublets $l_L^i$ and the right-handed singlets $l_R^i$ for leptons.
The bosonic part of the action is given by
\begin{eqnarray}
&&S_g=\frac{1}{2 g^2} \int d^4 x\ {\rm tr} F_{\mu \nu}  F_{\mu \nu},
\label{gaugeaction}
\\
&&S_h=\frac{1}{ \lambda}\int d^4 x\left\{\left(D_{\mu} H\right)^{\dagger}
\left(D_{\mu} H\right)+
\frac{1}{8}\left(H^{\dagger}H-v^2\right)^2\right\},
\label{higgsaction}
\end{eqnarray}
where $F_{\mu\nu}=\partial_{\mu} A_{\nu}-\partial_{\nu}
A_{\mu}-i\left[A_{\mu},A_{\nu}\right]$ and
$D_{\mu}=\partial_{\mu}-iA_{\mu}$.
The masses of the gauge boson and the Higgs boson are
\begin{eqnarray}
&&m_w=\sqrt{\frac{g^2}{2\lambda}} v,\quad m_h=\frac{1}{\sqrt{2}} v.
\end{eqnarray}
The fermionic part of the action is
\begin{eqnarray}
&&S_q=\int d^4 x (i q_L^{\dagger}\sigma_{\mu}D_{\mu}q_L
-i u_R^{\dagger}\bar{\sigma}_{\mu}\partial_{\mu} u_R
-i d_R^{\dagger}\bar{\sigma}_{\mu}\partial_{\mu} d_R\\
\nonumber
&& \hspace{10ex}+y_u q_L^{\dagger}\epsilon H^*
u_R-y_d q_L^{\dagger}H d_R +h.c.),\\
&&S_l=\int d^4 x (i l_L^{\dagger}\sigma_{\mu}D_{\mu}l_L
-i e_R^{\dagger}\bar{\sigma}_{\mu}\partial_{\mu} e_R
-y_l l_L^{\dagger} H e_R +h.c.),
\end{eqnarray}
where $\epsilon^{\alpha\beta}=-\epsilon_{\alpha\beta} = i \sigma^2$,
 $\sigma_{\mu}= (\sigma,i)$ and
$\bar{\sigma}_{\mu}= (\sigma,-i)$.
The masses of fermions are given by $m_{u,d,l}=y_{u,d,l}v$.
The total action is given by
\begin{eqnarray}
S=S_g+S_h+\sum_{i=1}^{n_f}\left(S_{l}+\sum_{a=1}^{3}S_{q}\right).
\end{eqnarray}

In order to calculate the Green function,
$\langle\psi^{4n_f}A_{\mu}^{n_w}H^{n_h}\rangle$, we must construct a
series of the configurations, which give a dominant contribution to
the path integral in the topologically non-trivial functional subspace.
They are not solutions of the equation of
motion.
The vacuum expectation value of the Higgs field breaks the
classical scale invariance and therefore, except the zero-radius instanton,
there exists no solution of the equation of motion in this model.

The dominating configurations make a valley of the action \cite{balyun,KR}.
Therefore, the proper valley method is suitable to construct them \cite{AKnv}.
The construction of this series of configurations is carried out in
Ref.\cite{AHSW}.
We dub these configurations valley instanton. The valley instanton is
the solution of a generalized equation of motion, the proper valley
equation which is called in Ref.\cite{AHSW} by the name of the new
valley equation.
The proper valley equation for this system is given by,
\begin{eqnarray}
&&\frac{\delta^2 S}{\delta A_{\mu}\delta A_{\nu}}F^A_{\nu}
+\frac{\delta^2 S}{\delta A_{\mu}\delta H^{\dagger}}F^H
+\frac{\delta^2 S}{\delta A_{\mu}\delta H}F^{H\dagger}
=\lambda_e F^A_{\mu}\nonumber,
\\
&&\frac{\delta^2 S}{\delta H^{\dagger}\delta A_{\mu}}
F^A_{\mu}+
\frac{\delta^2 S}{\delta H^{\dagger}\delta H}F^{H\dagger}
+\frac{\delta^2 S}{\delta H^{\dagger}\delta H^{\dagger}}F^H
=\lambda_e F^{H\dagger},
\label{valley}
\\
&&F^A_{\mu}=\frac{\delta S}{\delta A_{\mu}},
\quad F^H =\frac{\delta S}{\delta H},\nonumber
\end{eqnarray}
where the integration over the space-time is implicit.
The valley is parameterized by the eigenvalue $\lambda_e$ that
is identified with the zero mode corresponding to the scale invariance
at  $v\rightarrow 0$.

To simplify the equation, we adopt the following ansatz;
\begin{eqnarray}
A_{\mu}(x)=\frac{x_{\nu}\bar{\sigma}_{\mu\nu}}{x^2}\cdot 2a(r),
\quad H(x)=v\left( 1-h(r)\right)\eta,
\label{ansatz}
\end{eqnarray}
where $\eta$ is a constant isospinor, and $a$ and
$h$ are real dimensionless functions of dimensionless variable
$r$, which is defined by $r=\sqrt{x^2}/\rho$. The matrix
$\bar{\sigma}_{\mu\nu}$ is defined, according to the conventions of
Ref.\cite{Esp}, as $\bar{\sigma}_{\mu\nu}=\bar{\eta}_{a\mu\nu}\sigma^a/2$.
We have introduced the scaling parameter $\rho$. The tensor structure
in (\ref{ansatz}) is the same as that of the instanton  in the singular
gauge \cite{thooft}.

Inserting this ansatz to (\ref{valley}), the structure
of $F^A_{\mu}$ and $F^{H\dagger}$ is
determined as the following;
\begin{eqnarray}
F^A_{\mu}(x)=\frac{x_{\nu}\bar{\sigma}_{\mu\nu}}{x^2}\cdot
\frac{2v^2}{\lambda}f^a(r),
&&F^{H\dagger}(x)=-\frac{v^3}{\lambda}f^h(r)\eta.
\label{fansatz}
\end{eqnarray}

At $\rho v \rightarrow 0$, (\ref{valley}) reduces
to the equation of motion and the equation for the zero-mode
fluctuation around the instanton solution. The solution of this set of
equations is the following;
\begin{eqnarray}
\begin{array}{cc}
\displaystyle a_{0}=\frac{1}{1+r^{2}},
&\displaystyle h_{0}=1-\left(\frac{r^{2}}{1+r^{2}}\right)^{1/2},
\\[4mm]
\displaystyle f^a_{0}=\frac{2Cr^{2}}{(1+r^{2})^{2}},
&\displaystyle f^h_{0}=\frac{Cr}{(1+r^{2})^{3/2}},
\end{array}
\label{eq:a0}
\end{eqnarray}
where $C$ is an arbitrary function of $\rho v$.
Note that $a_{0}$ is the instanton solution in the singular gauge and
$h_{0}$ is the Higgs configuration in the instanton background \cite{thooft}.
We have adjusted the scaling parameter $\rho$ so that the radius of the
instanton solution is unity. The mode solutions
$f^a_{0}$ and $f^h_{0}$ are obtained from $\partial a_0/\partial \rho$ and
$\partial h_0/\partial \rho$, respectively.

We construct the valley instanton analytically for $\rho v \rsim 1$.
When $\rho v=0$, it is given by the ordinary instanton configuration $a_{0}$,
$h_{0}$, $f^{a}_{0}$ and $f^{h}_{0}$.
When $\rho v$ is small but not zero, it is expected that small $\rho v$
corrections appear in the solution.
On the other hand, at large distance from the core of the valley
instanton, this solution is expected to decay exponentially, because
gauge boson and Higgs boson are massive.
Therefore, the solution is similar to the
instanton near the origin and decays exponentially in the asymptotic region.
We solve the valley equation in both regions
and analyze the connection in the intermediate region.
In this manner we find the solution.

In the asymptotic region, $a$, $h$, $f^{a}$
and $f^{h}$ become small and the valley equation can be linearized.
The solution of equations is
\begin{eqnarray}
&&a(r)=C_{1}\,r\frac{d}{dr}G_{\rho m_w}(r)+\frac{1}{\nu}f^{a}(r),
\nonumber\\
&&h(r)=C_{2}\,G_{\rho m_h}(r)+\frac{1}{\nu}f^{h}(r),
\nonumber\\
&&\label{linear}\\
&&f^{a}(r)=C_{3}\,r\frac{d}{dr}G_{\rho \mu_w}(r),
\nonumber\\
&&f^{h}(r)=C_{4}\,G_{\rho \mu_h}(r),
\nonumber
\end{eqnarray}
where $C_{i}$ are arbitrary functions of $\rho v$, $\nu$ is defined as
$\lambda_e=v^2\nu/\lambda$ and $\mu_{w,h}$
are defined as
$\mu_{w,h}=m_{w,h}\sqrt{1-2\nu}$.
The function $G_{m}(r)$ is
\begin{equation}
G_{m}(r)=\frac{m K_{1}(m r)}{(2\pi)^{2}r},
\end{equation}
where $K_{1}$ is a modified Bessel function. As was expected above,
these solutions decay exponentially at infinity.
By matching two solutions
(\ref{eq:a0}) and (\ref{linear}) in the intermediated region, we
obtain $C_{1}+C_{3}/\nu=-2\pi^{2}$, $C_{2}+C_{4}/\nu=2\pi^{2}$,
$C_{3}=-4\pi^{2}C$ and $C_{4}=4\pi^{2}C $. By more detail analysis, we find
$C, \nu = 1/4$ at $\rho v \rightarrow 0 $ and we observe
$C,\nu \sim  1/4$ for $\rho v \rsim 1$ numerically \cite{AHSW}.
Inserting the valley instanton to the action,
we obtain,
\begin{equation}
\label{action}
S=\frac{8\pi^{2}}{g^{2}}
    +\frac{2\pi^{2}}{\lambda}(\rho v)^{2}+\cdots \,  .
\end{equation}

The fermionic zero mode around the valley instanton is given by,
\begin{eqnarray}
&&q_{Lr}^{\dot{\alpha}}(x)=-\frac{1}{\pi\rho^3}
x_{\nu}\bar{\sigma}_{\nu}^{\dot{\alpha} \beta}
\left(-\epsilon_{rs}\eta^{\dagger
s}\eta_{\beta}u_L(r)+\eta_r \epsilon_{\beta
\gamma}\eta^{\dagger\gamma}d_L(r)\right),\\
&&u_{R\alpha}(x)=\frac{i}{2\pi}
\frac{m_u}{\rho}u_R(r)\eta_{\alpha},\\
&&d_{R\alpha}(x)=\frac{i}{2
\pi}\frac{m_d}{\rho}d_R(r)\epsilon_{\alpha\gamma}
\eta^{\dagger\gamma},
\end{eqnarray}
where the Greek and the Roman letters denote indices of spinor and isospinor,
respectively.
For $\rho v \rsim 1$, the solution is given  approximately, by the
following;
\begin{eqnarray}
&&u_L(r)=\left\{
    \begin{array}{ll}
\displaystyle \frac{1}{r(r^2+1)^{3/2}},&
\quad \mbox{if} \quad r \ll (\rho m_u)^{-1/2}\ ; \\[0.4cm]
\displaystyle  -2 \pi^2 \frac{1}{r}\frac{d}{dr}G_{\rho m_u}(r),&
\quad \mbox{if} \quad r \gg (\rho m_u)^{-1/2}\ ;
     \end{array} \right.\\
&&u_R(r)=\left\{
    \begin{array}{ll}
\displaystyle \frac{1}{r^2+1},&
\quad \mbox{if} \quad r \ll (\rho m_u)^{-1/2}\ ; \\[0.4cm]
\displaystyle  2 \pi^2 G_{\rho m_u}(r),&
\quad \mbox{if} \quad r \gg (\rho m_u)^{-1/2},
     \end{array} \right.
\end{eqnarray}
and $d_L$ and $d_R$ are obtained by replacing $m_u$ with $m_d$.
The above approximative behavior is correct even for $\rho v \sim
1$, because the valley instanton keeps its shape unchanged for
$\rho v \sim 1$.
We obtain the zero
mode for leptonic sector in the similar manner.

\reseteqnum
\section{Collective coordinates of the valley instanton}
In this section, we give the collective coordinate method for the
valley instanton.
Because the valley instanton breaks the invariance under the translations
and the Lorentz $SU(2)_R$ rotations, we must introduce the collective
coordinates to restore it\footnote{In Higgs phase,
the collective coordinates for the global $SU(2)$ gauge rotations are
not needed as they are not symmetries.\cite{SV}}.
However, as the valley instanton is not a classical solution except
the zero radius one, the zero modes for the translations and the Lorentz
$SU(2)_R$ rotations do not exist in general.
Therefore the conventional collective coordinate method can not be used.
The basic strategy of our collective coordinate method is to treat the
translations and the Lorentz $SU(2)_R$ as a kind of ``gauge symmetry''
as well as the original gauge symmetry \cite{HK}.
This guarantees automatically the manifest invariance under the
translations and the Lorentz rotations through the following
deformation and do not require the zero modes.

In addition to these collective coordinates, the valley direction,
namely the $\rho$ integral, must be brought into.
This is because along the valley
direction the action of the system varies gently.
The corresponding
integration can not be approximated by the Gaussian integration and must
be performed more exactly.
The total number of the collective coordinates is eight (four collective
coordinates from the
translations,three from the Lorentz $SU(2)_R$ rotations and one from
the valley direction) and the
same as that of the BPST instanton.

First we rewrite the gauge field and the Higgs fields as
\noindent
\begin{eqnarray}
&&A_{\mu}(x)=A_{\mu}(x;\rho)+Q_{\mu}(x;\rho)\label{eq:col-gauge},\\
&&H(x)=H(x;\rho)+Q_{H}(x;\rho).\label{eq:col-Higgs}
\end{eqnarray}
Here $A_{\mu}(x;\rho)$ and $H(x;\rho)$ are the valley instanton and
$Q_{\mu}(x;\rho)$ and $Q_H(x;\rho)$ are quantum fluctuations around
them.
As was shown in (\ref{ansatz}),
the tensor structures of the valley instanton are
\begin{eqnarray}
&&A_{\mu}(x;\rho)=\frac{x_{\nu}\bar{\sigma}_{\mu\nu}}{x^2}\cdot 2a(r),\\
&&H(x;\rho)=v(1-h(r))\eta,
\end{eqnarray}
where $r$ is $r=\sqrt{x^2}/\rho$.
For the fermions, we consider them as the quantum fluctuations.

Substituting  (\ref{eq:col-gauge}) and (\ref{eq:col-Higgs}) for
the original Lagrangian, we obtain a Lagrangian for the quantum fluctuations.
Due to the invariance of the original Lagrangian, it
has the invariance under the gauge rotations, the translations and the
Lorentz $SU(2)_R$ rotations.
The transformation law of the quantum fluctuations are as follows.

\begin{enumerate}
\item  gauge rotations: $g(x)$
\begin{eqnarray}
&&u_{R}^g(x)=u_R(x),\quad u^{\dagger g}_{R}(x)=u_{R}^{\dagger}(x),
\nonumber\\
&&d_{R}^g(x)=d_{R}(x),\quad d_{R}^{\dagger g}(x)=d_R^{\dagger}(x),\nonumber\\
&&q_{L}^g(x)=g(x)q_{L}(x),
\quad q^{\dagger g}_{L}(x)=q^{\dagger}_{L}(x)g^{\dagger}(x),\\
&&Q_{\mu}^g(x;\rho)
=-i\partial_{\mu}g(x)g^{\dagger}(x)
+g(x)(A_{\mu}(x;\rho)+Q_{\mu}(x;\rho))g^{\dagger}(x)
-A_{\mu}(x;\rho),\nonumber\\
&&Q_{H}^{g}(x;\rho)=g(x)\left(H(x;\rho)+Q_{H}(x;\rho)\right)-H(x;\rho).
\nonumber
\end{eqnarray}

\item translations: $x_0$
\begin{eqnarray}
&&u_R^{x_0}(x)=u_R(x+x_0),
\quad u_R^{\dagger x_0}(x)=u^{\dagger}_R(x+x_0),\nonumber\\
&&d_R^{x_0}(x)=d_R(x+x_0),
\quad d_R^{\dagger x_0}(x)=d^{\dagger}_R(x+x_0),\nonumber\\
&&q_L^{x_0}(x)=q_L(x+x_0),
\quad q_{L}^{\dagger x_0}(x)=q_{L}^{\dagger}(x+x_0),\\
&&Q_{\mu}^{x_0}(x;\rho)
=Q_{\mu}(x+x_0;\rho)+A_{\mu}(x+x_0;\rho)-A_{\mu}(x;\rho),
\nonumber\\
&&Q_H^{x_0}(x;\rho)=Q_H(x+x_0;\rho)+H(x+x_0;\rho)-H(x;\rho).
\nonumber
\end{eqnarray}

\item Lorentz $SU(2)_R$ rotations: $U$
\begin{eqnarray}
&&u^U_{R\alpha}(x)=U_{\alpha}^{\beta}u_{R\beta}(R^{-1}x),\quad
u^{\dagger U}_{R\dot{\alpha}}(x)=u^{\dagger}_{R
\dot{\alpha}}(R^{-1}x),\nonumber\\
&&d^U_{R\alpha}(x)=U_{\alpha}^{\beta}d_{R\beta}(R^{-1}x),\quad
d^{\dagger U}_{R\dot{\alpha}}(x)=d^{\dagger}_{R
\dot{\alpha}}(R^{-1}x),\nonumber\\
&&q_L^{U\dot{\alpha}}(x)=q_L^{\dot{\alpha}}(R^{-1}x),\quad
q_L^{\dagger U\alpha }(x)
=q_L^{\dagger \beta}(R^{-1}x)U_{\beta}^{\dagger\alpha},\\
&&Q_{\mu}^{U}(x;\rho)
=\frac{x_{\nu}}{x^2}U\bar{\sigma}_{\mu\nu}U^{\dagger}\cdot 2a(r)
+R_{\mu\nu}Q_{\nu}(R^{-1}x;\rho)-A_{\mu}(x;\rho),\nonumber\\
&&Q_H^U(x;\rho)=Q_H(R^{-1}x;\rho),\nonumber
\end{eqnarray}
where $R^{-1}x=R^{-1}_{\mu\nu}x_{\nu}$ and the $SO(4)$ matrix $R$ is
defined as follows;
\begin{eqnarray}
&&R_{\mu\nu}\bar{\sigma}_{\nu}=\bar{\sigma}_{\mu}U,
\end{eqnarray}
and we have used the following equations.
\begin{eqnarray}
R_{\mu\nu}A_{\nu}(R^{-1}x;\rho)
=\frac{x_{\nu}}{x^2}
U^{\dagger}\bar{\sigma}_{\mu\nu}U\cdot 2a(r),\quad
H(R^{-1}x;\rho)=H(x;\rho).
\end{eqnarray}

\end{enumerate}
The transformation law of $l_{L}$ and $e_{R}$
is the same as that of $q_{L}$ and $d_{R}$ respectively.

We will treat these transformations as gauge symmetries.
To extract the ``gauge orbits''
from the path integral, we use the Faddeev-Popov technique.
For simplicity, we abbreviate the whole original
fields by the single filed $\Phi$, the valley instanton by $\Phi_0$
and the whole quantum fluctuations by $Q$.
We define the Faddeev-Popov determinant as
\begin{equation}
\Delta_{FP}(Q)^{-1}\equiv
\int {\cal D}gdUdx_0d\rho\,\delta\left(F(Q^{gx_0U}(x;\rho))\right)
\delta\left(\int d^4x\frac{1}{N(\rho)}\frac{\delta S}{\delta\Phi_{0}(x;\rho)}
Q^{gx_0U}(x;\rho)\right),
\label{eqn:FP}
\end{equation}
where $Q^{gx_0U}(x;\rho)$ is the quantum fluctuations transformed
successively under the gauge rotations $g(x)$, the translations $x_0$ and the
Lorentz $SU(2)_R$ rotations $U$, and
\begin{eqnarray}
&&\delta\left(F(Q^{gx_{0}U}(x;\rho))\right)
=\prod_x\delta\left(F_{\rm gauge}(Q^{gx_0U}(x;\rho))\right)
\prod_{\nu=1}^{4}\delta\left(F^{\nu}_{\rm trans}(Q^{gx_0U}(\rho))\right)
\nonumber\\
&&\hspace{28ex}\times
\prod_{a=1}^{3}\delta\left(F^{a}_{\rm Lorentz}(Q^{gx_0U}(\rho))\right),
\nonumber\\
&&N(\rho)=\int d^4x \left(\frac{\delta S}{\delta
\Phi_{0}(x;\rho)}\right)^2
\nonumber\\
&&\hspace{5.5ex}
=\int d^4x
\left[\frac{1}{2}\left(\frac{\delta S}{\delta A_{\mu}(x;\rho)}\right)^{2}
+\left(\frac{\delta S}{\delta H^{\dagger}(x;\rho)}\right)
\left(\frac{\delta S}{\delta H(x;\rho)}\right)
\right],\\
&&
\frac{\delta S}{\delta \Phi_{0}(x;\rho)}Q^{gx_{0}U}(x;\rho)
=\frac{\delta S}{\delta A_{\mu}(x;\rho)}Q_{\mu}^{gx_{0}U}(x;\rho)
+\frac{\delta S}{\delta H(x;\rho)}Q_{H}^{gx_{0}U}(x;\rho)
\nonumber\\
&&\hspace{28ex}
+\frac{\delta S}{\delta H^{\dagger}(x;\rho)}Q_{H}^{\dagger gx_0U}(x;\rho).
\nonumber
\end{eqnarray}
Here $F_{\rm gauge}$, $F_{\rm Lorentz}$ and $F_{\rm trans}$ are gauge
fixing functions, for example, which are given by
\begin{eqnarray}
&&F_{\rm gauge}(Q(x;\rho))=(D_{\mu}^cQ_{\mu})^a(x;\rho),\nonumber\\
&&F^{\nu}_{\rm trans}(Q(\rho))=\int d^4x\,{\rm tr}
\left\{f_{\mu}^{\nu}(x;\rho)Q_{\mu}(x;\rho)\right\}
,\label{eqn:gauge-fix}\\
&&F^a_{\rm Lorentz}(Q(\rho))=\int d^4x\,{\rm tr}
\left\{f^{a}_{\mu}(x;\rho)Q_{\mu}(x;\rho)\right\}
,\nonumber
\end{eqnarray}
where $f^{\nu}_{\mu}(x;\rho)$ and $f_{\mu}^{a}(x;\rho)$
are appropriate functions for the gauge-fixing.
The first delta-function in the right-hand side of (\ref{eqn:FP})
gives the gauge fixing and
the second one restricts the quantum fluctuation to be
orthogonal to the valley direction.

Now we consider the following Green function;
\begin{eqnarray}
&&\langle \psi(x_1)\cdots\psi(x_{4n_f})\,
A_{\mu_1}(y_1)\cdots A_{\mu_{n_w}}(y_{n_w})\,
\phi(z_1)\cdots\phi(z_{n_h})
\rangle\nonumber\\
&&={\cal N}
\int{\cal D}\Phi\, \psi(x_1)\cdots\psi(x_{4n_f})\,
A_{\mu_{1}}(y_1)\cdots A_{\mu_{n_w}}(y_{n_w})\,
\phi(z_1)\cdots\phi(z_{n_h})\,
e^{-S(\Phi)},
\end{eqnarray}
where ${\cal N}$ is an appropriate normalization factor and
$\psi(x)$ is the Dirac fermion corresponding to quark or lepton,
namely,
\begin{eqnarray}
\psi(x)=\left\{
\begin{array}{ll}
(u_{R}(x),q_{L\,1}(x))&\,\mbox{for quark in the up sector}\\
(d_{R},q_{L\,2}(x))&\,\mbox{for quark in the down sector}\\
(0,l_{L\,1}(x))&\,\mbox{for lepton in the up sector}\\
(e_{R}(x),l_{L\,2}(x))&\,\mbox{for lepton in the down sector}
\end{array}
\right. ,
\end{eqnarray}
and $\phi(x)$ is the shifted Higgs field, $\phi(x)=H(x)-v\eta$.
Inserting the following identity
\begin{equation}
1=\int {\cal D}gdUdx_0d\rho\,\Delta_{FP}(Q)
\delta\left(F(Q^{gx_0U}(x;\rho))\right)\delta
\left(\int d^4x\frac{1}{N(\rho)}\frac{\delta S}{\delta\Phi_{0}(x;\rho)}
Q^{gx_0U}(x;\rho)\right),
\end{equation}
we obtain
\begin{eqnarray}
&&\langle \psi(x_1)\cdots\psi(x_{4n_f})\,
A_{\mu_1}(y_1)\cdots A_{\mu_{n_w}}(y_{n_w})\,
\phi(z_1)\cdots\phi(z_{n_h})\rangle\nonumber\\
&&={\cal N}
\int {\cal D}gdUdx_0d\rho
\int{\cal D}\Phi
\Delta_{FP}(Q)\delta\left(F(Q^{gx_0U}(x;\rho))\right)\delta
\left(\int d^4x\frac{1}{N(\rho)}\frac{\delta S}{\delta\Phi_{0}(x;\rho)}
Q^{gx_0U}(x;\rho)\right)\nonumber\\
&&\hspace{5ex}\times \psi(x_1)\cdots\psi(x_{4n_f})\,
A_{\mu_1}(y_1)\cdots A_{\mu_{n_w}}(y_{n_w})
\,\phi(z_1)\cdots \phi(z_{n_h})\,
e^{-S(\Phi)}.
\end{eqnarray}
Substituting the inserted fields in the integral for
\begin{eqnarray}
&&u_{R\beta}(x_i)=U^{\dagger\delta}_{\beta}u_{R\delta}^{gx_0U}(R(x_i-x_0))
,\quad
u_{R\dot{\beta}}^{\dagger}(x_i)
=u_{R\dot{\beta}}^{\dagger gx_0U}(R(x_i-x_0)),
\nonumber\\
&&d_{R\beta}(x_i)=U^{\dagger\delta}_{\beta}d_{R\delta}^{gx_0U}(R(x_i-x_0))
,\quad
d_{R\dot{\beta}}^{\dagger}(x_i)
=d_{R\dot{\beta}}^{\dagger gx_0U}(R(x_i-x_0)),
\nonumber\\
&&q_{L}^{\dot{\alpha}}(x_i)
=g^{\dagger}(x_i)q_{L}^{gx_0U\dot{\alpha}}(R(x_i-x_0)),\quad
q_{L}^{\dagger\alpha}(x_i)
=q_{L}^{\dagger gx_0U\beta}(R(x_i-x_0))U^{\alpha}_{\beta}g(x_i),
\nonumber\\
&&e_{R\beta}(x_i)=U^{\dagger\delta}_{\beta}e_{R\delta}^{gx_0U}(R(x_i-x_0))
,\quad
e_{R\dot{\beta}}^{\dagger}(x_i)
=e_{R\dot{\beta}}^{\dagger gx_0U}(R(x_i-x_0)),\\
&&l_{L}^{\dot{\alpha}}(x_i)
=g^{\dagger}(x_i)l_{L}^{gx_0U\dot{\alpha}}(R(x_i-x_0)),\quad
l_{L}^{\dagger\alpha}(x_i)
=l_{L}^{\dagger gx_0U\beta}(R(x_i-x_0))U^{\alpha}_{\beta}g(x_i),
\nonumber\\
&&A_{\mu}(y_i)=\frac{(y_i-x_{0})_{\nu}}{(y_i-x_0)^2}
g^{\dagger}(y_i)U\bar{\sigma}_{\mu\nu}U^{\dagger}g(y_i)
\cdot 2a(r_{y_ix_0})
+ig^{\dagger}(y_i)\partial_{\mu}g(y_i)\nonumber\\
&&\hspace{7ex}+g^{\dagger}(y_i)R_{\mu\nu}^{-1}
Q_{\nu}^{gx_0U}(R(y_i-x_0);\rho)g(y_i),
\nonumber\\
&&\phi(z_i)=g^{\dagger}(z_i)v(1-h(r_{z_ix_0}))\eta-v\eta
+g^{\dagger}(z_i)Q_H^{gx_0U}(R(z_i-x_0);\rho),\nonumber
\end{eqnarray}
where $r_{y_ix_0}=\sqrt{(y_i-x_0)^2}/\rho$, and using the relations
\begin{eqnarray}
&&S(\Phi)=S(\Phi_0+Q^{gx_0U}),\nonumber\\
&&\Delta_{FP}(Q)=\Delta_{FP}(Q^{gx_0U}),\\
&&{\cal D}\Phi={\cal D}Q={\cal D}Q^{gx_0U},\nonumber
\end{eqnarray}
all the integration variables which appear in the path integral become
$Q^{gx_0U}$.
Then we can rename it bake to $Q$.
Furthermore, since the physical observables are
invariant under the $SU(2)$ gauge rotations,
we can simply set $g(x)=1$ in the path-integral instead of the
integration of $g(x)$.
This may breaks the translation and Lorentz invariance of the
Green functions which are not invariant under the $SU(2)$ gauge rotations,
however, the
lack of the explicit symmetries in such quantities is a gauge artifact.
For physical observables, the translation and Lorentz invariance are
not broken.
Finally, we obtain
\begin{eqnarray}
&&\langle \psi(x_1)\cdots\psi(x_{4n_f})\,
A_{\mu_1}(y_1)\cdots A_{\mu_{n_w}}(y_{n_w})\,\phi(z_1)\cdots
\phi(z_{n_h})
\rangle\nonumber\\
&&={\cal N}\int dUdx_0d\rho
\int{\cal D}Q
\Delta_{FP}(Q)\delta\left(F(Q(x;\rho))\right)\delta
\left(\int d^4x\frac{1}{N(\rho)}\frac{\delta S}{\delta\Phi_{0}(x;\rho)}
Q(x;\rho)\right)\nonumber\\
&&\hspace{5ex}\times \psi(x_1)\cdots\psi(x_{4n_f})\,
A_{\mu_1}(y_1)\cdots A_{\mu_{n_w}}(y_{n_w})
\,\phi(z_1)\cdots \phi(z_{n_h})\,
e^{-S(\Phi_0+Q)},
\end{eqnarray}
where the inserted fields in the integral are given by
\begin{eqnarray}
&&\psi(x_i)=
\left\{
\begin{array}{l}
\smallskip
\left(
\begin{array}{c}
U^{\dagger\beta}_{\alpha}u_{R\beta}(R(x_i-x_0))\\
q_{L1}^{\dot{\alpha}}(R(x_i-x_0))
\end{array}
\right)\\
\smallskip
\left(
\begin{array}{c}
U^{\dagger\beta}_{\alpha}d_{R\beta}(R(x_i-x_0))\\
q_{L2}^{\dot{\alpha}}(R(x_i-x_0))
\end{array}
\right)\\
\smallskip
\left(
\begin{array}{c}
0\\
l_{L1}^{\dot{\alpha}}(R(x_i-x_0))
\end{array}
\right)\\
\left(
\begin{array}{c}
U^{\dagger\beta}_{\alpha}e_{R\beta}(R(x_i-x_0))\\
l_{L2}^{\dot{\alpha}}(R(x_i-x_0))
\end{array}
\right)\\
\end{array}
\right.
,\quad
\psi^{\dagger}(x_i)=
\left\{
\begin{array}{l}
\smallskip
\left(
\begin{array}{c}
u^{\dagger}_{R\dot{\alpha}}(R(x_i-x_0))\\
q_{L1}^{\dagger\beta}(R(x_i-x_0))U^{\alpha}_{\beta}
\end{array}
\right)\\
\smallskip
\left(
\begin{array}{c}
d^{\dagger}_{R\dot{\alpha}}(R(x_i-x_0))\\
q_{L2}^{\dagger\beta}(R(x_i-x_0))U^{\alpha}_{\beta}
\end{array}
\right)\\
\smallskip
\left(
\begin{array}{c}
0\\
l_{L1}^{\dagger{\beta}}(R(x_i-x_0))U^{\alpha}_{\beta}
\end{array}
\right)\\
\left(
\begin{array}{c}
e^{\dagger}_{R\dot{\alpha}}(R(x_i-x_0))\\
l_{L2}^{\dagger\beta}(R(x_i-x_0))U^{\alpha}_{\beta}
\end{array}
\right)\\
\end{array}
\right.
,\nonumber\\
&&A_{\mu}(y_i)=\frac{(y_i-x_0)_{\nu}}{(y_i-x_0)^{2}}U
\bar{\sigma}_{\mu\nu}U^{\dagger}\cdot
2a(r_{y_ix_0})+R_{\mu\nu}^{-1}Q_{\nu}(R(y_i-x_0);\rho),
\nonumber\\
&&\phi(z_i)=-vh(r_{z_ix_0})\eta+Q_H(R(z_i-x_0);\rho)\nonumber.
\end{eqnarray}

\reseteqnum
\section{Calculation of the cross section}
In this section, we calculate the cross section for the process of the
type\  $\bar{\rm q}+\bar{\rm q}\rightarrow(3n_f-2){\rm q}+n_f
l+n_w {\rm W}+n_h {\rm H}$, using the valley instanton.
{}From Section 3, the Fourier transform of the Green
function is given by,
\begin{eqnarray}
\tilde{G}({p},{q},{k})&=& e^{-8\pi^2/g^2} g^{-8-n_w} \lambda^{-n_h/2}
\int dU\int d\rho e^{-2\pi^2(\rho v)^2/\lambda}\rho^{2 n_f-5} F(\rho
v,\rho \mu)
\nonumber
\\
&&\hspace{15ex} \times \tilde{\psi}(p_1) \cdots \tilde{A}_{\mu}(q_1)\cdots
\tilde{\phi}(k_1)\cdots ,
\label{grn1}
\end{eqnarray}
where  $\tilde{G}({p},{q},{k})$ is defined  by,
\begin{eqnarray}
&&(2\pi)^4\delta^4(p_1+\cdots+q_1+\cdots+k_1+\cdots)
\tilde{G}(p,q,k)\nonumber\\
&&\hspace{3ex}=\int d^4x_1\cdots d^4y_1\cdots d^4z_1\cdots
e^{ip_1\cdot x_1+\cdots+iq_1\cdot y_1+\cdots+ik_1\cdot z_1+\cdots}\nonumber\\
&&\hspace{20ex}\times\langle\psi(x_1)\cdots A(y_1)\cdots \phi(z_1)\cdots
\rangle.
\end{eqnarray}
The dimensionless function $F(\rho v, \rho \mu)$ includes the
contributions from the fermionic determinant, bosonic determinant and
jacobian factor:
\begin{eqnarray}
&&g^{-8}e^{-8\pi^2/g^2-2\pi^2(\rho v)^2/\lambda}
\rho^{2n_f-5}F(\rho v,\rho \mu)
\nonumber\\
&&=\frac{1}{Z_0}\int{\cal D}Q
\Delta_{FP}(Q)\delta\left(F(Q(x;\rho))\right)\delta
\left(\int d^4x\frac{1}{N(\rho)}\frac{\delta S}{\delta\Phi_{0}(x;\rho)}
Q(x;\rho)\right)e^{-S(\Phi_0+Q)}.
\end{eqnarray}
We introduce the renormalization point $\mu$, and $g$
and $\lambda$ in (\ref{grn1}) are renormalized coupling constants. In
the leading order, $\tilde{A}_{\mu}$ and $\tilde{\phi}$ in
(\ref{grn1}) are given by the valley
instanton and $\tilde{\psi}$ is given by the zero mode around the
valley instanton.
The Fourier transforms of the fermionic zero mode, $\tilde{\psi}(p)$
is given by,
\begin{eqnarray}
  \label{psitilde}
  \tilde{\psi}=
\left(
  \begin{array}{c}
\tilde{\psi}_{R\alpha}\\
\tilde{\psi}_L^{\dot{\alpha}}
  \end{array}
\right)
=2\pi i \rho
\left(
  \begin{array}{c}
-m_f U^{\dagger}\chi\\
\bar{\rlap/p}U^{\dagger}\chi
  \end{array}
\right)\frac{1}{p^2+m_f^2}+\cdots ,
\end{eqnarray}
where $\cdots$ includes the regular term in the limit:
$p^2+m_f^2\rightarrow 0$, and $\chi_{\alpha}$ is the constant spinor,
$(0,-1)$ for $\psi = u,\nu$ and (1,0) for $\psi = d,e$. We have used
the notation: $\bar{\rlap/p}=p_{\mu}\bar{\sigma}_{\mu} $. The Fourier
transforms of the valley instanton, $\tilde{A}_{\mu}$ is given by
\begin{eqnarray}
  \tilde{A}_{\mu} = -\frac{1}{\nu}\pi^2 i
\rho^2 \frac{q_{\nu}U \bar{\sigma}_{\mu\nu}U^{\dagger}}{q^2+m_w^2}+\cdots.
\end{eqnarray}
The Fourier transform of the shifted Higgs field $\phi$ is given by,
\begin{eqnarray}
  \label{phitilde}
  \tilde{\phi}=\frac{1}{2\nu}\pi^2 \rho^2 \frac{v}{k^2+m_h^2}\eta+\cdots .
\end{eqnarray}
By (\ref{grn1})-(\ref{phitilde}), we obtain
\begin{eqnarray}
\nonumber
\tilde{G}({p},{q},{k})&=&(-i)^{n_w}\cdot 2^{4 n_f-n_h}\pi^{4
  n_f+2n_w+2n_h} g^{-8-n_w}e^{-8\pi^2/g^2} v^{-6n_f-2n_w-n_h+4}\\
&&\hspace{15ex}\times D(\frac{\mu}{v};\lambda)\int dU L(U;p,q,k)
\label{grn}
\end{eqnarray}
where $D(\frac{\mu}{v};\lambda)$ and $L(U;p,q,k)$ are defined by,
\begin{eqnarray}
&&D(\frac{\mu}{v};\lambda)=\lambda^{-n_h/2}\int d(\rho v) e^{-2\pi^2
    (\rho v)^2/\lambda} (\rho v)^{6 n_f+2 n_w+2
      n_h-5}\nu^{-n_w-n_h}F(\rho v,\rho \mu),
\label{D}
\\
&&L(U;p,q,k)=\prod_{i=1}^{4n_f}
\left(
\begin{array}{c}
\displaystyle -m_f U^{\dagger}\chi
\\
\displaystyle
\bar{\rlap/p}_i U^{\dagger}\chi
\end{array}
\right)\frac{1}{p_i^2+m_f^2}
\prod^{n_w}_{j=1}\frac{q_{j\nu}U\bar{\sigma}_{\mu\nu}
U^{\dagger}}{q_j^2+m_{w}^2}
\prod^{n_h}_{l=1}\frac{\eta}{k_l^2+m_{h}^2}.
\end{eqnarray}
We ignore the momentum dependence that goes to zero on mass-shell.

Performing the LSZ procedure, we obtain the invariant amplitude
$T_{n_f,n_w,n_h}$. Summing up the polarization and charge
state of the gauge field, isospinor and spinor of the fermion field, we
obtain the amplitude;
\begin{eqnarray}
&&\sum |T_{n_f,n_w,n_h}|^2=2^{8n_f+n_w-2 n_h}\pi^{8n_f+4n_w+4n_h}
(g^2)^{-n_w-8}e^{-16\pi^2/g^2} v^{-2(6n_f+2n_w+n_h-4)}
\nonumber
\\
&&\hspace{10ex}\times D^2(\frac{\mu}{v};\lambda)\int dU\prod_{i=1}^{4
n_f}{\rm tr} (\bar{\rlap/p}_i
U)\prod_{j=1}^{n_w}(-)g^{\mu\rho}q^{\nu}_jq^{\sigma}_j
{\rm tr} (\sigma_{\mu\nu}U\bar{\sigma}_{\rho\sigma}U^{\dagger}).
\end{eqnarray}
The cross section is given by the following;
\begin{eqnarray}
&&\sigma=\frac{1}{n_w!n_h!}\frac{1}{2^4}\int
\prod_{i=3}^{4n_f}\frac{d^3 p_i}{(2\pi)^3 2 E_{p_i}}
\prod_{j=1}^{n_w}\frac{d^3 q_j}{(2\pi)^3 2 E_{q_j}}
\prod_{l=1}^{n_h}\frac{d^3 k_l}{(2\pi)^3 2 E_{k_l}}
(2\pi)^4\delta^{(4)}(p_{\rm in}-p_{\rm out})
\nonumber
\\
&&\hspace{15ex}\times \sum|T_{n_f,n_w,n_h}|^2\cdot \frac{1}{4\sqrt{(p_1\cdot
p_2)^2-m_1^2m_2^2} }
\nonumber
\\
&&\hspace{2ex}=\frac{1}{n_w!n_h!}2^{-4n_w-6n_h-5}(2\pi)^{-4
n_f+n_w+n_h+10}(g^2)^{-n_w-8} e^{-16\pi^2/g^2} v^{-2(6n_f+2n_w+n_h-4)}
\nonumber
\\
&&\hspace{10ex}\times
D^2(\frac{\mu}{v};\lambda)I_{2n_f,n_w,n_h}(s)\cdot s^{-1},
\end{eqnarray}
where $s=p_{\rm in}^2=(p_1+p_2)^2$ and the masses of fermions of the
initial state are ignored. We denote the sum of the momentums
of the final state by $p_{\rm out}$.
All the information about phase space and group integration is encoded
in the function $I_{2n_f,n_w,n_h}(s)$. We define $I_{l,m,n}(s)$ by,
\begin{eqnarray}
&&I_{l,m,n}(s)=
\int\prod_{i=3}^{2l}\frac{d^3 p_i}{2 E_{p_i}}
\prod_{j=1}^{m}\frac{d^3 q_j}{2 E_{q_j}}
\prod_{l=1}^{n}\frac{d^3 k_l}{2 E_{k_l}}
\delta^{(4)}(p_{\rm in}-p_{\rm out})
\nonumber
\\
&&\hspace{20ex}\times
2^{m}\int dU\prod_{i=1}^{2l}{\rm tr} (\bar{\rlap/p}_i U)
\prod_{j=1}^{m}(-)g^{\mu\rho}q^{\nu}_jq^{\sigma}_j {\rm tr}
(\sigma_{\mu\nu}U\bar{\sigma}_{\rho\sigma}U^{\dagger})
\nonumber
\\
&&\hspace{9ex}=\int\prod_{i=3}^{2l}\frac{d^3 p_i}{2 E_{p_i}}
\prod_{j=1}^{m}\frac{d^3 q_j}{2 E_{q_j}}
\prod_{l=1}^{n}\frac{d^3 k_l}{2 E_{k_l}}
\delta^{(4)}(p_{\rm in}-p_{\rm out})\nonumber
\\
&&\hspace{20ex}\times
\int dU\prod_{i=1}^{2l}{\rm tr}
(\bar{\rlap/p}_i U)\prod_{j=1}^{m}\left[\left\{{\rm
tr}(\bar{\rlap/q}_jU)\right\}^2-m_{w}^2\right]. \label{I}
\end{eqnarray}

To calculate the cross section, we must evaluate two functions,
$D(\mu/v;\lambda)$ and $I_{l,m,n}(s)$, which are given by (\ref{D})
and (\ref{I}), respectively. The former includes  $\rho$ integral and
the latter includes the phase space and group integral.

\subsection{Phase space and group integral}
We consider the Laplace transformation of $I_{l,m,n}(s)$\cite{BK}.
 At first, we assume that $p_{\rm in}$, $p_1$
and $p_2$ are independent variables, and at last we input $p_{\rm
in}=p_1+p_2$.
The Laplace transform $\Phi_{l,m,n}(\alpha;p_1,p_2)$ is given by,
\begin{eqnarray}
&&\Phi_{l,m,n}(\alpha;p_1,p_2)=\int d^4 p_{\rm in}\ e^{-\alpha\cdot
p_{\rm in}}I_{l,m,n}(p_{\rm in};p_1,p_2)
\nonumber
\\
&&\hspace{3ex}=\int dU{\rm tr}(\bar{\rlap/p}_1U){\rm
tr}(\bar{\rlap/p}_2U)
\phi_f(\alpha,U)^{2l-2}\phi_g(\alpha,U)^{m}\phi_h(\alpha)^{n},
\end{eqnarray}
where
\begin{eqnarray}
&&\phi_f(\alpha,U)=\int\frac{d^3 p}{2 E_p}{\rm
tr}(\bar{\rlap/p}U)e^{-\alpha\cdot p}
=2\pi \frac{m_f^2}{\alpha^2} K_2(\alpha m_f){\rm tr}(\bar{\rlap/\alpha}U),
\\
&&\phi_g(\alpha,U)=\int\frac{d^3 q}{2 E_q}\left[\left\{{\rm
tr}(\bar{\rlap/q}U)\right\}^2-m_{w}^2\right]e^{-\alpha\cdot q}
\nonumber
\\
&&\hspace{9ex}=\frac{2\pi m_{w}^3}{\alpha^3}K_3(\alpha
m_{w})\left[\left\{{\rm tr}(\bar{\rlap/\alpha}U)\right\}^2-\alpha^2\right],
\\
&&\phi_h(\alpha)=\int\frac{d^3 k}{2 E_k}e^{-\alpha\cdot k}=\frac{2\pi
m_{h}}{\alpha} K_1(\alpha m_{h}).
\end{eqnarray}
In the above,  $K_{\nu}$  is the modified Bessel function and
$\alpha=\sqrt{\alpha^2}$. Then we can perform the group
integration easily,
\begin{eqnarray}
&&\int dU {\rm tr}(\bar{\rlap/p}_1 U){\rm tr}(\bar{\rlap/p}_2 U)\left\{{\rm
tr}(\bar{\rlap/\alpha} U)\right\}^{2l-2}\left[\left\{{\rm
tr}(\bar{\rlap/\alpha}U)\right\}^2-\alpha^2\right]^{m}
\nonumber
\\
&&\hspace{5ex}=\sum_{i=0}^{m}\hspace{-1ex}\ _{m}C_i\int dU
{\rm tr}(\bar{\rlap/p}_1 U){\rm tr}(\bar{\rlap/p}_2 U)
\left\{{\rm tr}(\bar{\rlap/\alpha}
U)\right\}^{2l-2+2i}(-\alpha^2)^{n-i}
\nonumber
\\
&&\hspace{5ex}=\left\{C_0(p_1\cdot\alpha)(p_2\cdot\alpha)+
C_1(p_1\cdot p_2)\alpha^2 \right\}\alpha^{2(l+m-2)},
\end{eqnarray}
where
\begin{eqnarray}
&&C_0=4\sum_{i=0}^{m}(-)^{m-i}\hspace{-1ex}\ _{m}C_i
\frac{\left\{2(l+i-1)\right\}!}{(l+i+1)!(l+i-2)!},
\\
&&C_1=2\sum_{i=0}^{m}(-)^{m-i}\hspace{-1ex}\ _{m}C_i
\frac{\left\{2(l+i-1)\right\}!}{(l+i+1)!(l+i-1)!}.
\end{eqnarray}
Finally, using the inverse Laplace transformation, we obtain the
function $I_{l,m,n}$. We can evaluate $I_{\l,m,n}$ analytically in the
case of the extremely relativistic limit and the non-relativistic limit.
In the extremely relativistic limit, $I_{\l,m,n}$ is given by,
\begin{eqnarray}
I_{l,m,n}=C^{\rm ER}_{l,m,n}s^{3l+2m+n-4},
\end{eqnarray}
where
\begin{eqnarray}
&&C^{\rm
ER}_{l,m,n}=2^m\left(\frac{\pi}{2}\right)^{2l+m+n-3}
\left\{(3l+2m+n-3)!(3l+2m+n-5)!\right\}^{-1}
\nonumber
\\
&&\hspace{7ex}\times\sum_{i=0}^{m}(-)^{m-i}\ _{m}C_i
\frac{\left\{2(l+i-1)\right\}!}{(l+i+1)!(l+i-1)!}
\left\{1+(l+i)(3l+2m+n-4)\right\}.
\end{eqnarray}
In the non-relativistic limit, we obtain
\begin{eqnarray}
I_{l,m,n}=C^{\rm
NR}_{l,m,n}m_f^{3/2}m_{w}^{5m/2}m_{h}^{n/2}(\sqrt{s}-M)^{6l+3(m+n)/2-10}
\cdot
s^{1/4},
\end{eqnarray}
where
\begin{eqnarray}
&&C^{\rm
NR}_{l,m,n}=\left\{\left(6l+\frac{3}{2}m
+\frac{3}{2}n-10\right)!\right\}^{-1}\cdot
2^{4l+(m+n)/2-7}\pi^{2l+3(m+n)/2-3}
\nonumber
\\
&&\hspace{10ex}\times \sum_{i=0}^m(-)^{m-i}\
_mC_i\frac{\left\{2(l+i-1)\right\}!}{(l+i+1)!(l+i-1)!}(l+i).
\end{eqnarray}
We assume the fermions are massless except for top quark and define
$M=m_f+n_w m_w +n_h m_h$.
In the general cases, we can perform the inverse Laplace transformation
numerically, using the steepest descent approximation.

\subsection{$\rho$ integral}
We evaluate the function $D(\mu/v;\lambda)$ defined by (\ref{D}). In
(\ref{D}), $F$ and $\nu$ depends on $\rho v$, non-trivially. However,
according to Ref.\cite{AHSW}, the valley instanton is nearly equal to BPST
instanton in the region  $\rho v \rsim 1$ and therefore we approximate
the forms
of $F$ and $\nu$ by that of BPST instanton, if $\rho$ integral is
dominated by the region $\rho v \rsim 1$.

For $\rho v \rsim 1$,
\begin{eqnarray}
  \nu\sim\frac{1}{4},\hspace{4ex} F(\rho v,\rho \mu)\sim c (\rho
  \mu)^{(43-8 n_f)/6},
\label{nuF}
\end{eqnarray}
where $c$ is a numerical constant \cite{thooft}
\begin{eqnarray}
  c\sim 861.94 e^{-0.997 n_f}2^{10 +  6n_f}\pi^{6+4n_f}.
\end{eqnarray}
We can calculate $\rho$ integral by (\ref{nuF}) and obtain
\begin{eqnarray}
  D(\frac{\mu}{v};\lambda)=2^{2n_f+2n_h-1}c
  \lambda^{-n_h/2}\left(\frac{\lambda}{2\pi^2}\right)^t
  \left(\frac{\mu}{v}\right)^{(43-8n_f)/6} \Gamma(t),
\end{eqnarray}
where $t=7 n_f/3+n_w+n_h+19/12$.
The integral is dominated by the contribution around the saddle
point;
\begin{eqnarray}
\rho v_s=\left\{\frac{\lambda}{2\pi^2}(t-\frac{1}{2})\right\}^{1/2}.
\end{eqnarray}
Therefore, we can check the consistency of the approximation by this
value of the saddle point. The saddle point $\rho v_s$ depends on the
number of bosons monotonously. The valley instanton is quite similar to
the original instanton even at $\rho v \sim  1$, and therefore the
approximation is valid for $n_w+n_h\sim 40$, where we assume
$m_{h}=m_{w}$ and use $g^2\sim 0.42$. On the other hand, the
constrained instanton is deviated from the original instanton at $\rho
v= 0.5$ \cite{AHSW} and so the approximation breaks for $n_w +n_h\sim 4$.
Therefore we can calculate the amplitude of the multi-boson process,
using the valley instanton.
Finally, we obtain
\begin{eqnarray}
&&\sigma=\frac{c^2}{n_w!n_h!}2^{-n_h-7}(2\pi)^{-40
  n_f/3-3n_w-3n_h+11/3}\Gamma(t)^2
\nonumber
\\
&&\hspace{5ex}\times\left(g^2\right)^{14
  n_f/3+n_w+n_h-29/6}e^{-16\pi^2/g^2}
m_w^{-2(6n_f+2n_w+n_h-4)}\left(\frac{\mu}{m_w}\right)^{(43-8n_f)/3}.
\end{eqnarray}
\begin{figure}
\centerline{
\epsfxsize=13cm
\epsfbox{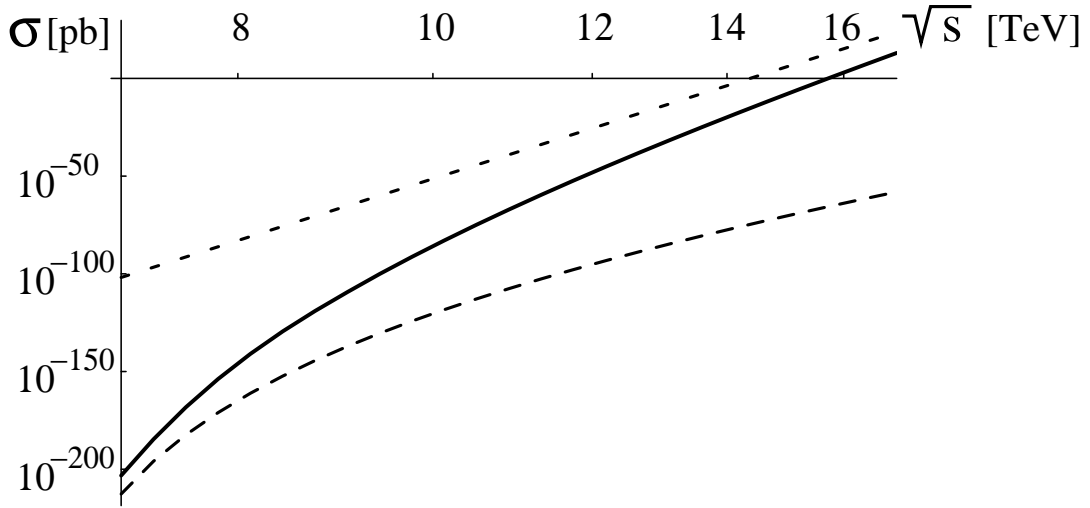}
}
\caption{The energy dependence of the cross section for $n_w=75$ and
$n_h=0$. The solid line denotes the result by the saddle point
approximation. The dashed line and the dotted line denote the
approximate result in the non-relativistic limit and extremely
relativistic limit, respectively. We set $n_f=3$, $\mu=m_w=80 {\rm GeV}$,
$g^2=0.42$, $m_f=180 {\rm GeV}$ and $m_h=200{\rm GeV}$ and neglect the other
fermion masses.}
\label{crs75}
\end{figure}

\begin{figure}
\centerline{
\epsfxsize=13cm
\epsfbox{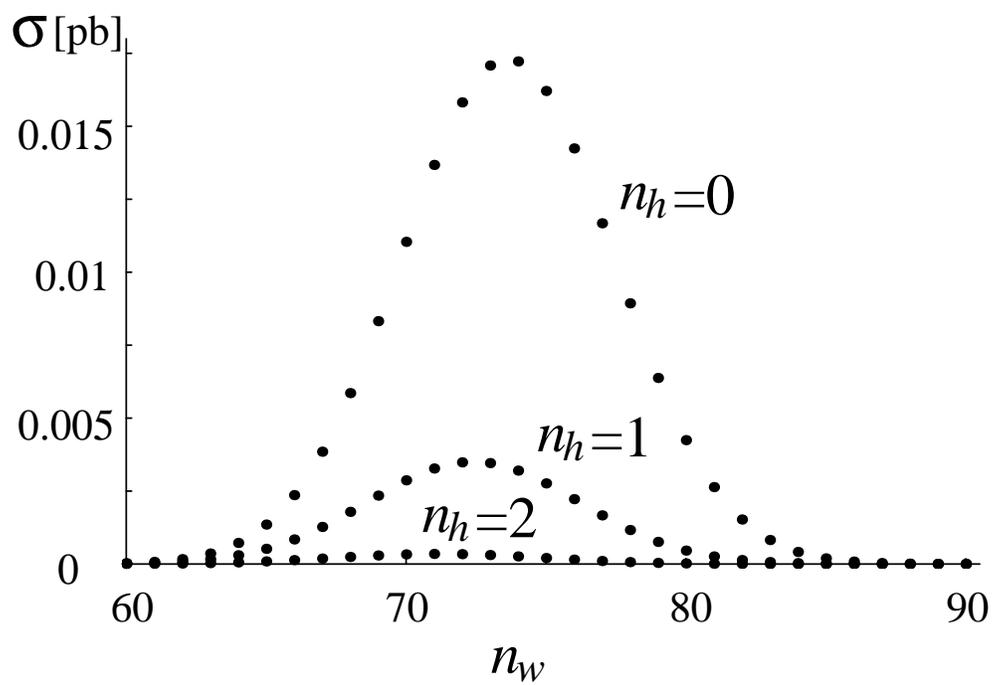}
}
\caption{The cross section for the various $n_w$ and $n_h$ at
$s^{1/2}=$ 15.55 TeV.
}
\label{crs-nw}
\end{figure}

\begin{figure}
\centerline{
\epsfxsize=13cm
\epsfbox{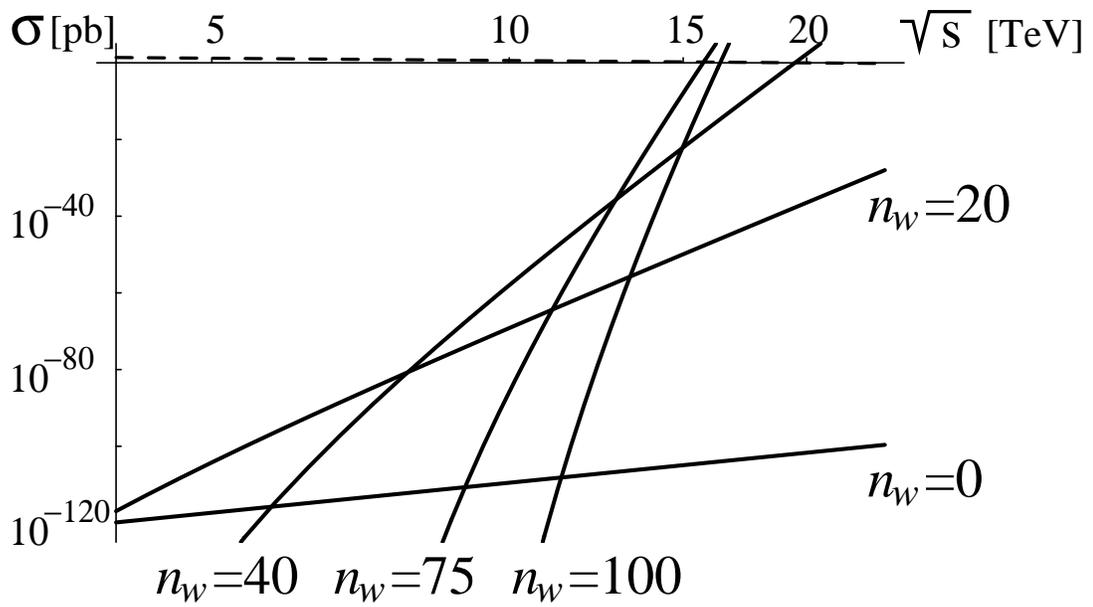}
}
\caption{The energy dependence of the cross section for $n_w$=0, 20, 40,
75 and 100, and $n_h$=0. A dashed line denotes the unitarity bound.}
\label{crs-s}
\end{figure}

\subsection{Results}
We show the
result in Fig.\ref{crs75} for the case where $n_w=75$ and $n_h=0$. In
Fig.\ref{crs75}, the solid line denotes the numerical result by the
steepest descent method. The dashed line and the dotted line denote the
approximate result in the non-relativistic limit and extremely
relativistic limit, respectively. We show cross
sections for the various values of $n_w$ and $n_h$ in Fig.\ref{crs-nw}
at $\sqrt{s}$=15.55 TeV. From this, we understand that the gauge boson
plays an important role rather than the Higgs particle. In
Fig.\ref{crs-s}, we show the energy dependence of the cross section
for $n_w$=0, 20, 40, 75 and 100, and $n_h$=0. We show the unitarity
bound as a dashed line in Fig.\ref{crs-s}. The approximation breaks in
the region of the energy
where the cross section overcomes the unitarity bound. In the region,
an interaction between the gauge bosons in
the final state \cite{RS} and interaction between multi-instantons
\cite{AK} play important roles, which are not evaluated in our analysis.
\reseteqnum
\section{Conclusion}
In this paper, we have applied the valley instanton to the baryon and
lepton number violating processes.
The constrained instanton method was used in the past works, but
the processes with many bosons, which are expected to be observed, are
out of the validity of the method.
By using the proper valley method, we have given
a reliable calculation of the process with many bosons.
First, by treating the Lorentz rotations and the translations as the gauge
symmetries, we have introduced the collective coordinates for the valley
instanton which is not a classical solution.
Then we have calculated the cross section of the baryon and
lepton number violating processes.
Furthermore, we have taken into account the instanton
orientation effect, boson and top quark mass effect, which are not
considered yet.
Then we have obtained more accurate result than earlier works. The final result
matches the qualitative estimate \cite{RS}. The cross section grows
up, as total energy becomes higher, and the processes with many gauge
bosons and no Higgs particle grow up more rapidly.

For high-energy processes associated with many
bosons, we see the unitarity is broken,
which means that our analysis is not appropriate in such cases.
It is because we don't take into account the multi-instanton effects\cite{AK}
or the interaction among the gauge bosons.
By the proper valley method, new phenomena might be revealed in the
subjects.
One of the reason is that there appear new terms in the interaction
between the valley instanton and the anti valley instanton.
Our present study is the first step toward these subjects.

\vskip1cm
\centerline{\large\bf Acknowledgment}
\noindent
We thank H.~Aoyama, H.~Kikuchi and S.~Wada for the numerous valuable
discussions.
T.~H.'s work is supported in part by the Grant-in-Aid for JSPS
fellows.

\appendix
\vskip 2cm
\centerline{\Large\bf Appendix}
\reseteqnum
\section{Faddeev-Popov determinant}

If we choose the gauge conditions as (\ref{eqn:gauge-fix}) and ignore
higher loop corrections, the Faddeev-Popov determinant becomes
\begin{eqnarray}
&&\Delta_{FP}(\rho)=8\pi^2\cdot{\rm Det}
\left|
\begin{array}{cccc}
A^{ab}(x,y;\rho)&B^{a\nu}(x;\rho)&C^{ab}(x;\rho)&0\\
D^{\mu b}(y;\rho)&E^{\mu\nu}(\rho)&F^{\mu b}(\rho)&X^{\mu}(\rho)\\
G^{ab}(y;\rho)&H^{a\nu}(\rho)&I^{ab}(\rho)&Y^{a}(\rho)\\
0&0&0&W(\rho)
\end{array}
\right|\nonumber\\
&&\hspace{8ex}=
8\pi^2\cdot{\rm Det}\left|
\begin{array}{ccc}
A^{ab}(x,y;\rho)&B^{a\nu}(x;\rho)&C^{ab}(x;\rho)\\
D^{\mu b}(y;\rho)&E^{\mu\nu}(\rho)&F^{\mu b}(\rho)\\
G^{ab}(y;\rho)&H^{a\nu}(\rho)&I^{ab}(\rho)\\
\end{array}
\right|
\times |W(\rho)|,
\end{eqnarray}
where
\begin{eqnarray}
&&A^{ab}(x,y;\rho)=\left(D_{\lambda}^0D_{\lambda}^0\right)_{ab}\delta(x-y)
\nonumber\\
&&\hspace{11.5ex}=\left\{\partial_{\lambda}^2\delta_{ab}
+2\epsilon_{acb}A_{\lambda}^c(x;\rho)\partial_{\lambda}
+\epsilon_{acd}\epsilon_{deb}A_{\lambda}^c(x;\rho)A_{\lambda}^e(x;\rho)
\right\}\delta(x-y),\nonumber\\
&&B^{a\nu}(x;\rho)=-2{\rm tr}
\left\{\left[i\frac{\sigma^a}{2},A_{\lambda}(x;\rho)\right]
\partial_{\nu}A_{\lambda}(x;\rho)\right\},
\nonumber\\
&&C^{ab}(x;\rho)=-2{\rm tr}\left\{
\left[i\frac{\sigma^a}{2},A_{\lambda}(x;\rho)\right]
\left[i\frac{\sigma^b}{2},A_{\lambda}(x;\rho)\right]
\right\},
\nonumber\\
&&D^{\mu b}(y;\rho)=-{\rm tr}\left\{
D_{\lambda}^0f_{\lambda}^{\mu}(y;\rho)\frac{\sigma^b}{2}\right\},
\nonumber\\
&&E^{\mu\nu}(\rho)=\int d^4x\,{\rm tr}
\left\{f_{\lambda}^{\mu}(x;\rho)\partial_{\nu}A_{\lambda}(x;\rho)\right\}
,\nonumber\\
&&F^{\mu b}(\rho)=\int d^4x\,{\rm tr}\left\{f_{\lambda}^{\mu}(x;\rho)
\left[i\frac{\sigma^b}{2},A_{\lambda}(x;\rho)\right]\right\},
\nonumber\\
&&G^{ab}(y;\rho)=-{\rm tr}
\left\{D_{\lambda}^0f_{\lambda}^a(y;\rho)\frac{\sigma^b}{2}\right\},\\
&&H^{a\nu}(\rho)=\int d^4x \,
{\rm tr}\left\{f_{\lambda}^{a}(x;\rho)\partial_{\nu}
A_{\lambda}(x;\rho)\right\}
,\nonumber\\
&&I^{ab}(\rho)=\int d^4x\, {\rm tr}
\left\{f_{\lambda}^{a}(x;\rho)
\left[i\frac{\sigma^b}{2},A_{\lambda}(x;\rho)\right]\right\},\nonumber\\
&&X^{\mu}(\rho)=-\int d^4x\,{\rm tr}
\left\{f_{\lambda}^{\mu}(x;\rho)
\frac{\partial A_{\lambda}(x;\rho)}{\partial \rho}
\right\},\nonumber\\
&&Y^a(\rho)=-\int d^4 x\,{\rm tr}\left\{f_{\lambda}^{a}(x;\rho)
\frac{\partial A_{\lambda}(x;\rho)}{\partial \rho}\right\},\nonumber\\
&&W(\rho)=-\frac{1}{N(\rho)}\int d^4x\,
\left[\frac{\delta S}{\delta A_{\lambda}^a(x;\rho)}
\frac{\partial A_{\lambda}^a(x;\rho)}{\partial \rho}
+\frac{\delta S}{\delta H(x;\rho)}
\frac{\partial H(x;\rho)}{\partial \rho}
+\frac{\partial H^{\dagger}(x;\rho)}{\partial \rho}
\frac{\delta S}{\delta H^{\dagger}(x;\rho)}\right].\nonumber\\\nonumber
\end{eqnarray}

When $\rho v\rightarrow 0$, the instanton measure
given by our collective coordinate method is the same as one given
by 't Hooft\cite{thooft}.
To see this, it is convenient to take $f_{\mu}^{\nu}(x;\rho)$ and
$f_{\mu}^{a}(x;\rho)$ as
\begin{eqnarray}
&&f_{\mu}^{\nu}=\partial_{\nu}A_{\mu}(x;\rho)-D_{\mu}^0A_{\nu}(x;\rho),\\
&&f_{\mu}^{a}(x;\rho)
=\left[i\frac{\sigma^a}{2},A_{\mu}(x;\rho)\right]
-D_{\mu}^0\left[\frac{\sigma^{a}}{2}\frac{\rho^2}{x^2+\rho^2}\right],
\end{eqnarray}
where $D_{\mu}^0$ is covariant derivative under the background of
the valley instanton; $D_{\mu}^0*=\partial_{\mu}*-i[A_{\mu}(x;\rho),*]$.
Then it holds
\begin{eqnarray}
&&E^{\mu\nu}(\rho)=\int d^4 x\,
{\rm tr}\left\{f_{\lambda}^{\mu}(x;\rho)
f_{\lambda}^{\nu}(x;\rho)\right\}
+\int d^4x\,{\rm tr}
\left\{f_{\lambda}^{\mu}(x;\rho)D_{\lambda}^0A_{\nu}(x;\rho)\right\},\\
&&F^{\mu b}(\rho)=\int d^4x\,
{\rm tr}\left\{f_{\lambda}^{\mu}(x;\rho)f_{\lambda}^{b}(x;\rho)\right\}
+\int d^4x\,
{\rm tr}\left\{f_{\lambda}^{\mu}(x;\rho)D_{\lambda}^0
\left[\frac{\sigma^b}{2}\frac{\rho^2}{x^2+\rho^2}\right]
\right\},\\
&&H^{a\nu}(\rho)=\int d^4 x\,{\rm tr}\left\{f_{\lambda}^{a}(x;\rho)
f_{\lambda}^{\nu}(x;\rho)\right\}
+\int d^4x\,{\rm tr}\left\{f_{\lambda}^a(x;\rho)D_{\lambda}^0
A_{\nu}(x;\rho)\right\},\\
&&I^{ab}(\rho)
=\int d^4x\,{\rm tr}\left\{f_{\lambda}^{a}(x;\rho)
f_{\lambda}^{b}(x;\rho)\right\}
+\int d^4x\,
{\rm tr}\left\{f_{\lambda}^{a}(x;\rho)D_{\lambda}^0
\left[\frac{\sigma^b}{2}\frac{\rho^2}{x^2+\rho^2}\right]
\right\}.
\end{eqnarray}
When $\rho v\rightarrow 0$,
the valley instanton becomes the BPST instanton and
$f_{\mu}^{\nu}$ and $f_{\mu}^{a}$ become the zero modes corresponding to
the global gauge rotations and the translations, which satisfy
\begin{eqnarray}
D_{\mu}^0f_{\mu}^{\nu}=0,\quad D_{\mu}^0f_{\mu}^{a}=0.
\end{eqnarray}
Therefore, as $\rho v\rightarrow 0$, we obtain
\begin{eqnarray}
&&D^{\mu b}(y;\rho)\rightarrow 0,\quad
G^{ab}(y;\rho)\rightarrow 0,\nonumber\\
&&E^{\mu\nu}(\rho)\rightarrow\int d^4x\,{\rm tr}
\left\{f_{\lambda}^{\mu}(x;\rho)f_{\lambda}^{\nu}(x;\rho)\right\}
=\delta_{\mu\nu}\Vert f_{\lambda}^{\mu}\Vert^2\cdot\frac{1}{2},\nonumber\\
&&F^{\mu a}(\rho)\rightarrow\int d^4x\,{\rm tr}
\left\{f_{\lambda}^{\mu}(x;\rho)f_{\lambda}^{b}(x;\rho)\right\}=0,\\
&&H^{a\nu}(\rho)\rightarrow\int d^4x\,{\rm tr}
\left\{f_{\lambda}^{a}(x;\rho)f_{\lambda}^{\nu}(x;\rho)\right\}=0,\nonumber\\
&&I^{ab}(\rho)\rightarrow
\int d^4x\,{\rm tr}\left
\{f_{\lambda}^{a}(x;\rho)f_{\lambda}^{b}(x;\rho)
\right\}=\delta_{ab}\Vert f_{\lambda}^{a}\Vert^2\cdot\frac{1}{2}.\nonumber
\end{eqnarray}
{}From (\ref{fansatz}) and (\ref{eq:a0}), it holds
\begin{eqnarray}
\frac{\delta S}{\delta A_{\mu}(x;\rho)}=F_{\mu}^A\rightarrow
\frac{\rho v^2}{4\lambda}\frac{\partial A_{\mu}(x;\rho)}{\partial \rho},
\quad \frac{\delta S}{\delta H(x;\rho)}=F^{H}\rightarrow
\frac{\rho v^2}{4\lambda}\frac{\partial H(x;\rho)}{\partial \rho},
\end{eqnarray}
as $\rho v\rightarrow 0$ and
\begin{eqnarray}
\int d^4x \left|\frac{\partial H(x;\rho)}{\partial\rho}\right|^2
=O(\rho v)\int d^4x\,
\left(\frac{\partial A_{\mu}(x;\rho)}{\partial \rho}\right)^2.
\end{eqnarray}
Then
\begin{eqnarray}
\left|W(\rho)\right|\rightarrow \sqrt{\int d^4x\,
\left(
\frac{\partial A_{\mu}^a(x;\rho)}{\partial\rho}\right)^2}
=\left\Vert\frac{\partial A_{\mu}}{\partial \rho}\right\Vert.
\end{eqnarray}
Therefore, we obtain
\begin{eqnarray}
\Delta_{FP}(\rho)\rightarrow
8\pi^2\cdot2^{-7}\cdot\left|{\rm Det}A^{ab}(x,y;\rho)\right|\cdot
\left\Vert\frac{\partial A_{\mu}}{\partial\rho}\right\Vert\cdot
\Vert f_{\mu}^{\nu}\Vert^8\cdot\Vert f_{\mu}^{a}\Vert^6,
\end{eqnarray}
as $\rho v\rightarrow 0$.
When we integrate the quantum fluctuations, the following factor
appears;
\begin{eqnarray}
\left(\frac{1}{\sqrt{2\pi}}\right)^{4}
\cdot\left(\frac{2}{\Vert f_{\mu}^{\nu}\Vert}\right)^4
\cdot\left(\frac{2}{\Vert f_{\mu}^a\Vert}\right)^3.
\end{eqnarray}
Then the Jacobian becomes
\begin{eqnarray}
8\pi^2\cdot\left(\frac{1}{\sqrt{2\pi}}\right)^{4}
\cdot\left\Vert\frac{\partial A_{\mu}}{\partial\rho}\right\Vert\cdot
\Vert f_{\mu}^{\nu}\Vert^4\cdot\Vert f_{\mu}^{a}\Vert^3.
\end{eqnarray}
This coincide with one given by 't Hooft,
since
\begin{eqnarray}
\left\Vert\frac{\partial A_{\mu}}{\partial\rho}\right\Vert
\rightarrow 4\pi,\quad
\Vert f_{\mu}^{\nu}\Vert\rightarrow 2\sqrt{2}\pi,\quad
\Vert f_{\mu}^a\Vert\rightarrow 2\pi\rho,
\end{eqnarray}
when $\rho v\rightarrow 0$.
Because the determinant parts coming from the path
integral agree with that of BPST instanton when $\rho v=0$, the
instanton measure of the valley instanton coincides that of BPST
instanton in this limit.

\newcommand{\J}[4]{{\sl #1} {\bf #2} (19#3) #4}
\newcommand{\MPL}{Mod.~Phys.~Lett.}
\newcommand{\NP}{Nucl.~Phys.}
\newcommand{\PL}{Phys.~Lett.}
\newcommand{\PR}{Phys.~Rev.}
\newcommand{\PRL}{Phys.~Rev.~Lett.}
\newcommand{\AP}{Ann.~Phys.}
\newcommand{\CMP}{Commun.~Math.~Phys.}
\newcommand{\CQG}{Class.~Quant.~Grav.}
\newcommand{\PRP}{Phys.~Rept.}
\newcommand{\SPU}{Sov.~Phys.~Usp.}
\newcommand{\RMPA}{Rev.~Math.~Pur.~et~Appl.}
\newcommand{\SPJ}{Sov.~Phys.~JETP}
\newcommand{\MP}{Int.~Mod.~Phys.}

\end{document}